\documentclass[prd, twocolumn, nofootinbib]{revtex4}

\usepackage{epsfig} 

\newcommand{\beq}{\begin{equation}}
\newcommand{\eeq}{\end{equation}}
\newcommand{\beqa}{\begin{eqnarray}}
\newcommand{\eeqa}{\end{eqnarray}}
\newcommand{\om}{\Omega_M}

\def\hmpcinv{h\,{\rm Mpc}^{-1}}

\def\fun#1#2{\lower3.6pt\vbox{\baselineskip0pt\lineskip.9pt
  \ialign{$\mathsurround=0pt#1\hfil##\hfil$\crcr#2\crcr\sim\crcr}}}

\begin{document} 

\title{Separating Dark Physics from Physical Darkness: \\ 
Minimalist Modified Gravity vs.\ Dark Energy} 
\author{Dragan Huterer$^1$ and Eric V.\ Linder$^2$} 
\affiliation{$^1$Kavli Institute for Cosmological Physics and 
Astronomy and Astrophysics Department, University of Chicago, 
Chicago, IL 60637 \\ 
$^2$Berkeley Lab, University of California, Berkeley, CA 94720}

\begin{abstract} 
The acceleration of the cosmic expansion may be due to a new component of
physical energy density or a modification of physics itself.  Mapping the
expansion of cosmic scales and the growth of large scale structure in tandem
can provide insights to distinguish between the two origins.  Using Minimal
Modified Gravity (MMG) -- a single parameter gravitational growth index
formalism to parameterize modified gravity theories -- we examine the
constraints that cosmological data can place on the nature of the new physics.
For next generation measurements combining weak lensing, supernovae distances,
and the cosmic microwave background we can extend the reach of physics to allow
for fitting gravity simultaneously with the expansion equation of state,
diluting the equation of state estimation by less than 25\% relative to when
general relativity is assumed, and determining the growth index to 8\%.
For weak lensing we examine the level of understanding needed of 
quasi- and nonlinear structure formation in modified gravity theories, 
and the trade off between stronger precision but greater susceptibility 
to bias as progressively more nonlinear information is used. 

\end{abstract} 


\date{\today}

\maketitle

\section{Introduction} \label{sec:intro}

Whether the acceleration of the cosmic expansion is due to a new 
physical component or a modification of gravitation, the answer will 
involve groundbreaking new physics beyond the current standard models 
for high energy physics and cosmology.  To obtain the clearest, unbiased 
picture of the fundamental physics we need to allow for the possibility 
of gravitation beyond Einstein and see where the data lead.  
This article presents simultaneous fitting of the cosmic expansion and 
the theory of gravity. 

Direct measurements of the expansion history can be interpreted generally 
as the equation of state of the universe; this may or may not correspond 
to a physical component.  Measurements of the growth 
history of mass fluctuations combine information on the expansion and 
the theory of gravity.  The two in complementarity thus allow separation 
of the possible origins for cosmic acceleration -- a physical dark 
component, e.g.\ a new field in high energy physics, or dark (new) physics, 
e.g.\ a modification of Einstein gravity. 

Within a specific theory of modified gravity one can attempt to calculate the
cosmological observables and determine the goodness of fit with data.  This
model dependent approach must proceed on a case by case basis and moreover
suffers from difficulties in computation of many quantities due to the
complexity of the theories.  Nor are the modifications necessarily well
motivated or completely free from pathologies.  The alternate approach taken
here is phenomenological, using a model independent yet physically reasonable
and broad parameterization of the gravity modification to gain insight into the
effect of generalizing Einstein gravity.  This is closely analogous to the
widely successful equation of state approach to 
modifications of the expansion history, giving model independent constraints
and understanding.

In \S\ref{sec:index} we discuss the development of the parametrized 
approach and its range of validity.  \S\ref{sec:data} lays out the 
cosmological probes, fiducial models, and survey data characteristics 
used in our analysis.  
We examine in \S\ref{sec:constrain} 
the ability of next generation cosmological probes to reveal new gravity 
simultaneous with fitting cosmological expansion, 
and the cosmological bias incurred if we neglect to 
allow the possibility of beyond Einstein gravity.  
Leverage and systematics from the nonlinear regime are discussed in 
\S\ref{sec:mods}. 

\section{Gravitation and Growth} \label{sec:index} 

Attempting to invent a general prescription for taking into account 
the effects of modified gravity is like seeking a general treatment of 
non-Gaussianity: there are so many ways in which a theory can be ``not'', 
in which a symmetry can be broken, that it seems a hopeless task.  However, 
we do not seek an all-encompassing description of modified gravity in 
all its aspects, but rather its gross effects on cosmological observables 
beyond the expansion rate.  

As a beginning step, we take a fairly conservative approach we call 
Minimal Modified Gravity (MMG).  The modifications
we consider are small, since general relativity gives a predominantly
successful description of the universe and its large scale structure, and
homogeneous, i.e.\ not dependent on environment \`a la chameleon scenarios
\cite{chameleon}.  We assume that structure formation continues to be described
adequately by growth from Gaussian density perturbations.  For weak
gravitational lensing, we do not have to get too deeply involved in the
nonlinear density regime and can concentrate on the growth law; we assume
changes to the gravitational deflection law are negligible.

This approach seeks to build understanding of modified gravity by 
taking a modest step away from general relativity.  By examining an 
alteration in the linear perturbation growth, preserving the standard 
mapping from linear to nonlinear fluctuations, we obtain a clear 
picture of a specific effect of modified gravity (see \S\ref{sec:mods} 
for relaxation of the assumption of standard mapping and discussion of 
scale dependence).  This serves as a proxy for presumably more complicated 
and model dependent effects. 

The form of the growth equation can be written so as to directly show 
the influence of the cosmic expansion and the additional effect of the 
gravity theory.  For a matter density perturbation $\delta\equiv\delta\rho 
/\rho$, the linear growth factor $g=\delta/a$ (scaling out the matter 
dominated universe behavior $\delta\sim a$) evolves in general relativity as 
\beqa 
g''&+&\left[5+\frac{1}{2}\frac{d\ln H^2}{d\ln a}\right]\,a^{-1}g' \nonumber \\ 
&+&\left[3+\frac{1}{2}\frac{d\ln H^2}{d\ln a}-\frac{3}{2}\om(a)\right] 
a^{-2}g=0,
\label{eq:g_2ndorder} 
\eeqa 
where a prime denotes derivative with respect to the scale factor $a$, 
$H=\dot a/a$ is the Hubble parameter, and $\om(a)$ is the dimensionless 
matter density.  Since the global cosmology parameters $H$ and $\om(a)$ 
are, essentially, the expansion history, we see that the cosmic 
expansion determines the structure growth. 

To make the relation between growth probes and expansion probes such as 
the luminosity distance-redshift relationship $r_l(z)$ even more explicit, 
we can start from the growth equation as 
\beq 
\ddot\delta+2H\dot\delta-4\pi\rho\delta=0, \label{eq:ddotdelta}
\eeq 
where a dot denotes time derivative, and use 
\beq 
r_l=a^{-1}\int dt/a = a^{-1}\int da/(a^2H) 
\eeq 
(for a flat universe to keep the notation simple) to write 
\beq 
\frac{d^2\delta}{dr_l^2}(a^{-2}-Hr_l)^2-\frac{d\delta}{dr_l}(Ha^{-2}+
\ddot a/a)-4\pi\rho\delta=0. 
\eeq 
Thus in general relativity the growth clearly contains the same 
cosmological information as the distance relation. 

As discussed in \cite{groexp}, we can alter the growth equation in modified
gravity by changing the matter source term (proportional to $\om(a)$ in 
Eq.~(\ref{eq:g_2ndorder}) or $\rho\delta$ in Eq.~(\ref{eq:ddotdelta})) 
or adding 
a new source through a nonzero right hand side.  Green function solutions for
such modifications are given in \cite{groexp}.  The matter source term can be
written as $Q\delta$, and $Q=\nabla^2\Phi/\delta$ arises from the equivalent of
the Poisson equation relating the metric potential $\Phi$ to the matter
perturbation $\delta$.  One possibility is to make a phenomenological
modification to the Poisson equation and investigate its effects on structure
growth; this has been investigated by 
\cite{whitekoch,sealfon,shirata,stabenaujain}
and we revisit it in \S\ref{sec:mods}.  An intriguing approach of general 
quadrature relations between the matter perturbations and metric potentials 
is discussed by \cite{bertschinger}.

A key aspect to note is that much of the growth {\it is\/} determined by the
expansion history, even in modified gravity, so we should not throw away that
knowledge.  By following the effects, treating the expansion in terms of the
well-developed equation of state formalism (whether arising from a physical
dark energy or a modified Friedmann equation), i.e.\
\beqa
H^2(z)/H_0^2&=&\om(1+z)^3+\delta H^2(z) \\
w(z)&=&-1+\frac{1}{3}\frac{d\ln \delta H^2}{d\ln (1+z)},
\eeqa
and adding a new parameter to
incorporate the effects of modified gravity specifically on the growth source
term, we render the physics appropriately.  This was the motivation behind the
gravitational growth index formalism of \cite{groexp}, which we follow,
calling the ansatz MMG.
An alternate approach is to define wholly separate growth variables
(see, e.g., \cite{knoxsongtyson,ishak}).

The gravitational growth index serves as a proxy for the full modified gravity
theory.  The linear growth factor is approximated by
\beq 
g(a)= e^{\int_0^a d\ln a\,[\om(a)^\gamma-1]}, 
\label{eq:groexp}
\eeq 
where $\gamma$ is the growth index.  This was shown to be accurate to 0.2\%
compared to the exact solution within general relativity for a wide variety of
physical dark energy equation of state ratios.  We verify explicitly that for
dynamical dark energy where the equation of state ratio is parametrized as
$w(a)=w_0+w_a(1-a)$, the formulas for $\gamma(w)$ given in \cite{groexp}
recover $g(a)$ to better than 0.3\% when $w_0+w_a<-0.1$.  That is, the growth
index parametrization of linear growth is extremely robust as long as the early
matter dominated epoch is not upset.

The gravitational growth index formalism has also been tested and found
accurate to 0.2\% for a single modified gravity scenario \cite{groexp}, DGP
\cite{dgp,luess} braneworld gravity, giving $\gamma=0.68$.  
We conjecture that it may work for
modified gravity theories with monotonic, smooth (Hubble timescale) evolution
in the source term, so long as the matter dominated epoch is not disrupted.  
For example, in DGP the source term receives a smooth
correction $1-(1/3)(1-\om^2(a))/(1+\om^2(a))$
\cite{luess,koyama_maartens,sawicki_song_hu,song_sawicki_hu}.  
Preliminary results in scalar-tensor theory also indicate successful 
approximation \cite{bmpl}.  Future work
includes testing this for other specific models; here we use the growth index
as an indicator of possible effects of modified gravity, with the advantage of
knowing at least it is robust for many cases beyond $\Lambda$CDM.

\section{Fiducial Model and Cosmological Data} \label{sec:data} 

To assess the leverage of cosmological observations to reveal dark 
physics vs.\ physical darkness, we 
simultaneously fit nine cosmological parameters: $A$, the normalization of
the primordial power spectrum at $k_{\rm fid}=0.05\hmpcinv$; physical matter
and baryon densities $\Omega_M h^2$ and $\Omega_B h^2$, spectral index $n$, sum
of the neutrino masses $m_{\nu}$, matter energy density today relative
to critical $\Omega_M$, and parameters describing the effective dark
energy equation of state $w_0$ and $w_a$, where $w(a)=w_0+(1-a)w_a$.  The mass
power spectrum $\Delta^2(k, a)\equiv k^3 P(k, a)/(2\pi^2)$ is written as 

\begin{equation}
\Delta^2(k, a) = \frac{4A}{25\om^2} 
\left ({k\over k_{\rm fid}}\right )^{n-1}
\left ({k\over H_0  }\right )^4
g^2(a)\,T^2(k)\,T_{\rm nl}(k, a)
\end{equation}

\noindent where $T(k)$ is the 
transfer function, and $T_{\rm nl}(k, a)$ is the prescription for the nonlinear
power.  Modified gravity enters through the ninth parameter, the gravitational
growth index $\gamma$ in the linear growth function $g(a)$ from
Eq.~(\ref{eq:groexp}), with the growth index $\gamma=0.55$ for the fiducial
cosmology corresponding to a flat universe with Einstein gravity and a
cosmological constant.  The other fiducial values adopted correspond roughly to
the current concordance cosmology, with $\Omega_M=0.3$, $w_0=-1$, $w_a=0$,
$\Omega_B h^2=0.023$, $\Omega_M h^2=0.14$, $n=0.97$, $m_{\nu}=0.2$ eV (one
massive species), and $A=2\times 10^{-9}$ (corresponding to $\sigma_8\simeq
0.9$).  While the exact values of some of these parameters, especially
$\sigma_8$, are still a subject of much debate, we do not expect that different
values allowed by current data will change any of our conclusions on the
detectability of non-standard growth.

The linear power spectrum uses the fitting formulae of \cite{Hu_transfer}.  
We always use the linear growth function from Eq.~(\ref{eq:groexp}) and account
for its dependence on cosmological parameters $\Omega_M$, $w_0$, $w_a$ and
$\gamma$ when taking the derivatives for the Fisher matrix.  To complete the
calculation of the full nonlinear power spectrum we use the halo model 
fitting formulae of \cite{smith}.

For the cosmological probes, we assume future weak lensing (WL) and Type Ia 
supernova (SN) data as
provided by the SNAP experiment \cite{SNAP} as well as cosmic microwave
background anisotropy (CMB) data provided by the upcoming Planck satellite 
\cite{planck}.

In this work, for weak lensing we only consider the two point correlation 
function.
The weak lensing shear power spectrum measures cosmology through both the 
mass power spectrum and distance factors, 
\begin{equation}
P_{ij}^{\kappa}(\ell) = \int_0^{\infty} \frac{dz}{r(z)^2 H(z)} 
\,W_i(z)\,W_j(z) \,
 \Delta^2\! \left ({\ell\over r(z)}, z\right ),
\label{eq:pk_l}
\end{equation} 
\noindent where $r(z)$ is the comoving distance and $H(z)$ is the Hubble
parameter.  The weights $W_i$ are given by $W_i(\chi) = (3/2)\,\Omega_M\,
H_0^2\,f_i(\chi)\, (1+z)$ where 
$f_i(\chi) = r(\chi)\int_{\chi}^{\infty} d\chi_s
n_i(\chi_s)\, r(\chi_s-\chi)/r(\chi_s)$, 
$\chi$ is the radial coordinate and $n_i$ is
the comoving density of galaxies if $\chi_s$ falls in the distance range
bounded by the $i$th redshift bin and zero otherwise.  
We employ the redshift
distribution of galaxies of the form $n(z)\propto z^2\exp(-z/z_0)$ that peaks
at $2z_0=1.0$.  The observed convergence power spectrum is \cite{tomography}
\begin{equation}
C^{\kappa}_{ij}(\ell)=P_{ij}^{\kappa}(\ell) + 
\delta_{ij} {\langle \gamma_{\rm int}^2\rangle \over \bar{n}_i},
\label{eq:C_obs}
\end{equation}
\noindent where $\langle\gamma_{\rm int}^2\rangle^{1/2}$ is the rms intrinsic
shear in each component, taken to be $0.22$, 
and $\bar{n}_i$ is the average number of galaxies in the $i$th
redshift bin per steradian.  The cosmological constraints can then be computed
from the Fisher matrix
\begin{equation}
F^{\rm WL}_{ij} = \sum_{\ell} \,{\partial {\bf C}\over \partial p_i}\,
{\bf Cov}^{-1}\,
{\partial {\bf C}\over \partial p_j},\label{eq:latter_F}
\end{equation}
\noindent where $p_i$ are the cosmological parameters and ${\bf Cov}^{-1}$ is
the inverse of the covariance matrix between the observed power spectra whose
elements are given by
\begin{eqnarray}
{\rm Cov}\left [C^{\kappa}_{ij}(\ell), C^{\kappa}_{kl}(\ell')\right ] &=& 
{\delta_{\ell \ell'}\over (2\ell+1)\,f_{\rm sky}\,\Delta \ell} \times \\
&&\left [ C^{\kappa}_{ik}(\ell) C^{\kappa}_{jl}(\ell) + 
  C^{\kappa}_{il}(\ell) C^{\kappa}_{jk}(\ell)\right ]. \nonumber 
\label{eq:Cov}
\end{eqnarray}

\noindent The fiducial WL survey assumes 1000 square degrees with tomographic 
measurements in 10 uniformly wide redshift bins extending out to $z=3$. The 
effective source galaxy density is 100 per square arcminute.  

We will also sometimes consider a South Pole Telescope (SPT \cite{SPT}) type
cluster survey with sky coverage of 5000 deg$^2$ and a total of about 25000
clusters (for the assumed cosmology with $\sigma_8=0.9$), giving a
number-redshift test involving the geometric volume and the number density from
growth of structure.  For simplicity, we neither consider the additional
information provided by masses of the clusters, nor degradation of constraints
due to the imperfectly calibrated mass-observable relation (here the observable
is the Sunyaev-Zel'dovich flux).  Our previous tests have shown that these two
effects roughly cancel out in the final cosmological constraints
\cite{cluster_photoz}.  Then the cluster Fisher matrix is
\begin{equation}
F^{\rm clus}_{ij}=\sum_{k} {1\over N(z_k)}
{\partial N(z_k)\over \partial p_i}
{\partial N(z_k)\over \partial p_j}
\label{eq:clus_fisher}
\end{equation}
\noindent where $N(z_k)$ is number of clusters in $k$th redshift bin.

The SN survey provides a luminosity distance-redshift test, with 2800 SNe
distributed in redshift out to $z=1.7$ as given by \cite{SNAP}, and combined
with 300 local supernovae uniformly distributed in the $z=0.03-0.08$ range. We
add systematic errors in quadrature with intrinsic random Gaussian errors 
of 0.15 mag per SNe.  The systematic errors create an effective error floor 
of $0.02\,(1+z_i)/2.7$ mag per bin of $\Delta z=0.1$ centered at redshift
$z_i$.  

For the CMB we use
the full Fisher matrix predicted for the Planck experiment with polarization
information (W.\ Hu, private communication).  Note that most, though not all,
information about dark energy is captured in the distance to the last
scattering surface from the acoustic peaks of the power spectrum (e.g.\
\cite{frieman}).  The effective precision of the angular diameter distance to
$z=1089$ from Planck is 0.4\% with temperature and polarization information
\cite{EisHuTeg}.

\section{Fitting Gravity} \label{sec:constrain} 

The combination of probes of the expansion history of the universe 
and the growth history of large scale structure tests the nature of 
the acceleration physics.  The expansion history is described by 
the effective equation of state parameters $w_0$ and $w_a$, and 
the deviation of the growth history from that given by Einstein gravity 
under that expansion is measured by the growth index $\gamma$.  Almost 
all cosmological analyses to date, however, have assumed Einstein gravity 
or worked within a specific alternate theory of gravity, rather than 
fitting for gravity. 

Ignoring the possibility of modified gravity creates the risk of the biasing
our cosmological conclusions.  This holds not only for ``gravitational''
parameters but all information.  Neglecting possible modification is equivalent
to fixing the gravitational growth index $\gamma$ to its Einstein value (e.g.\
$\gamma=0.55$ for general relativity and a cosmological constant model);
however this will bias the other parameters due to their covariances with
$\gamma$ (this ``gravity's bias'' was illustrated for the linear growth factor
alone, rather than the weak lensing shear power spectrum, in Fig.~5 of
\cite{bias}).

Suppose the true
value of the growth index differs by $\Delta\gamma$ from its assumed general
relativity value.  This propagates through into the weak lensing shear 
cross power spectrum
$C_\alpha^\kappa(\ell)$ at multipole $\ell$ for the pair of redshift bins
$\alpha\equiv \{i, j\}$, changing it from the assumed general relativity value
of $\bar C_\alpha^\kappa(\ell)$.  Using the Fisher matrix formalism, the bias
on any of the $P$ cosmological parameters is

\begin{eqnarray}
\delta p_i &=& \tilde{F}_{ij}^{-1} \sum_{\ell}
\left [C_{\alpha}^{\kappa}(\ell)-\bar C_{\alpha}^{\kappa}(\ell)\right ] \nonumber\\
&&\times \,{\rm Cov}^{-1}\left [\bar{C}^{\kappa}_{\alpha}(\ell), 
  \bar{C}^{\kappa}_{\beta}(\ell)\right ]
\,{\partial \bar C_{\beta}^{\kappa}(\ell) \over \partial p_j}\\[0.2cm]
&\approx& (\Delta\gamma) \tilde{F}_{ij}^{-1} F_{jg}
\label{eq:bias}
\end{eqnarray}

\noindent where the last line follows if the finite difference is replaced by a
derivative.  Here $\tilde{F}$ is the $(P-1)\times (P-1)$ Fisher matrix that
specifically does not include the growth index $\gamma$, $F$ is the full
$P\times P$ Fisher matrix, summations over $j$ and the redshift bin indices
$\alpha$, $\beta$ are implied, and $g$ is the index corresponding to
the $\gamma$ parameter in the matrix $F$.

\begin{figure}[!t]
\begin{center} 
\psfig{file=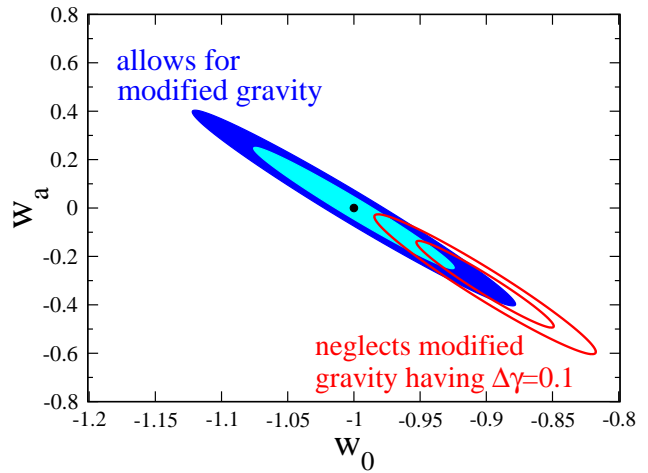, height=3.5in, angle=-90} 
\caption{68\% and 95\% CL constraints on the expansion history equation of
state parameters $w_0$ and $w_a$, marginalized over all other parameters. The
two blue (filled) contours give constraints from combining weak lensing,
supernovae, cluster, and CMB data, while simultaneously fitting for beyond
Einstein gravity.  The black dot shows the fiducial model. The two red (empty)
contours show the biased constraints if the modified gravity growth rate, 
here with $\Delta\gamma=0.1$, is ignored (i.e.\ fixing $\gamma$ to the 
general relativity value).  }
\label{fig:w0wa} 
\end{center} 
\end{figure}

Bias in the cosmology from neglecting the possibility of modified gravity can
be significant.  The red (open) contours in Fig.~\ref{fig:w0wa} demonstrate
that assuming Einstein gravity in a universe with $\gamma$ actually higher by
0.1 can shift the expansion characteristics (effective equation of state
parameters) by $\sim2\sigma$.  Recall that the DGP braneworld model has
$\Delta\gamma=0.13$ with respect to the standard cosmological constant case.
Note too that within this simple treatment of modified gravity a shift in
$\gamma$ moves the $w_0$-$w_a$ contour along the degeneracy direction, else the
bias would be even larger.  Given the price of closing our eyes to the issue of
possible gravitational modifications, we must attempt to fit for beyond
Einstein deviations.

Including the gravitational growth index as an additional parameter, and
marginalizing over it to estimate the effective equation of state parameters
describing the expansion, removes the bias but necessarily degrades parameter
determination.  (This would of course become more severe if we required more
growth variables than just $\gamma$.)

The degradation on the weak lensing shear constraints from marginalizing 
over $\gamma$ causes a factor $\sim2$ increase in the contour area.  While 
adding CMB or supernovae data does not directly constrain the growth, 
they prove valuable in breaking degeneracies between parameters.  
Adding CMB to WL improves constraints by 30-35\%, but still suffers 
the factor of 2 weakening relative to fixing $\gamma$.  The inclusion of 
SN is potent in reducing uncertainties.  WL+SN+CMB allows for fitting 
modified gravity, improving parameter estimation by 5-7 times and the area 
constraint by 40 times over WL alone.  Conversely, it only dilutes 
estimation of $w_0$ by 23\%, $w_a$ by 14\%, and the contour area by 
35\% relative to the ``fixed to Einstein'' case.  This seems 
a fairly modest price to pay for extending the physics reach to beyond 
Einstein gravity. 

The marginalized uncertainties on the key parameters describing the nature of
the effective dark physics are shown in Table~\ref{tab:constraints}.  Since SN
and CMB do not probe growth, we start with WL measurements and add other probes
in sequence.

\begin{table}[!t]
  \caption{Fiducial constraints on the gravitational growth index 
$\gamma$ (last column), 
as well as the two expansion history parameters $w_0$ and $w_a$. 
Starting with the weak lensing survey, we then add the supernova, CMB, and 
cluster information consecutively. Note that both WL and cluster surveys
are partially sensitive to nonlinear physics; constraints that only rely
on the linear regime are discussed in \S\ref{sec:mods}.
}
  \label{tab:constraints}
  \begin{tabular}[t]{|l|c|c||c|}
    \hline
    Probe & $\sigma(w_0)$ & $\sigma(w_a)$ & $\sigma(\gamma)$ \\
    \hline\hline  
    Weak lensing    & 0.33  & 1.16  & 0.23\\\hline 
    + SNe Ia        & 0.06  & 0.28  & 0.10\\\hline 
    + Planck        & 0.06  & 0.21  & 0.044\\\hline 
    + Clusters      & 0.05  & 0.16  & 0.037\\\hline 
  \end{tabular}
\end{table}

\begin{figure}[!th]
\begin{center} 
\psfig{file=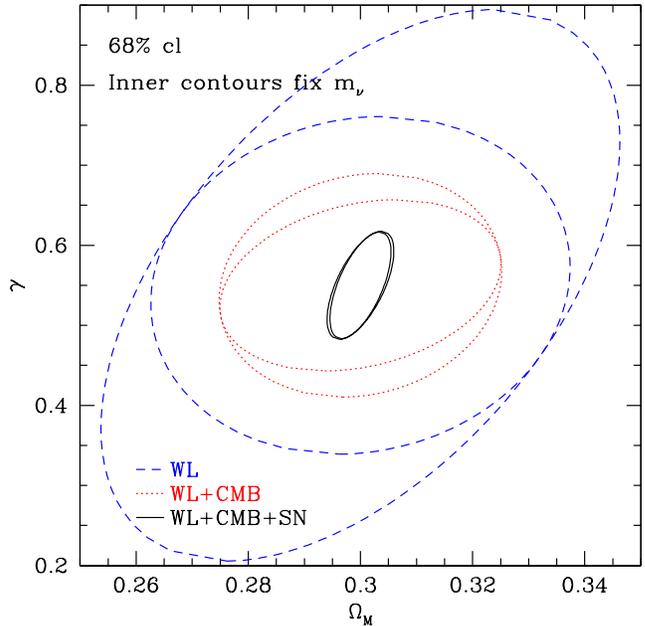,width=3.5in} 
\caption{68\% CL constraints on the gravitational growth index $\gamma$ 
and the matter density $\om$, marginalizing over the effective equation 
of state (and all other parameters).  To fit for beyond Einstein gravity 
as well as the expansion history requires a combination of probes.  
Other effects on growth, such as neutrino masses, should be taken into 
account as well; the inner contour of each pair shows the effect of 
holding this fixed, most severe for weak lensing in isolation. 
}
\label{fig:og} 
\end{center} 
\end{figure}

While marginalizing over the growth index allows us to account for 
modified gravity rather than ignoring it, we also want to measure the 
modification itself.  That is, we want to extract information 
quantifying the deviation, to guide us toward any ``dark physics'', 
not just say that there is some inconsistency with general relativity. 
Figure~\ref{fig:og} show the constraints on the growth index from the 
cosmological probes, in the $\gamma$-$\om$ plane (marginalizing over 
the equation of state and other parameters). 

We see an enormous difference in the leverage on modified gravity as 
we combine probes, again due to the breaking of degeneracies.  The 
determination of $\gamma$ reaches 0.044 for the combination WL+SN+CMB, 
representing 8\% precision with respect to the Einstein value.  If 
we do not include supernova data, then the uncertainty increases by 
a factor 2 (5 with only WL), and furthermore the overall contour area 
in the $\gamma$-$\om$ plane increases by 11 times (44 for WL only).   
So complementarity between WL and SN is quite important for answering 
the question of whether we are facing dark physics or physical darkness. 

Complementarity of probes is important for separation of different 
physical effects on growth.  While inclusion of neutrino mass, which 
can suppress growth (as does increasing $\gamma$), 
broadens the uncertainty area of WL by 74\%, adding other methods 
immunizes against such a ``theory systematic'', with the effect on 
WL+SN+CMB limited to 10\%. 

\section{Structure in the Nonlinear Regime}\label{sec:mods}

\subsection{Beyond the linear regime} 

\begin{figure*}[!t]
\begin{center} 
\psfig{file=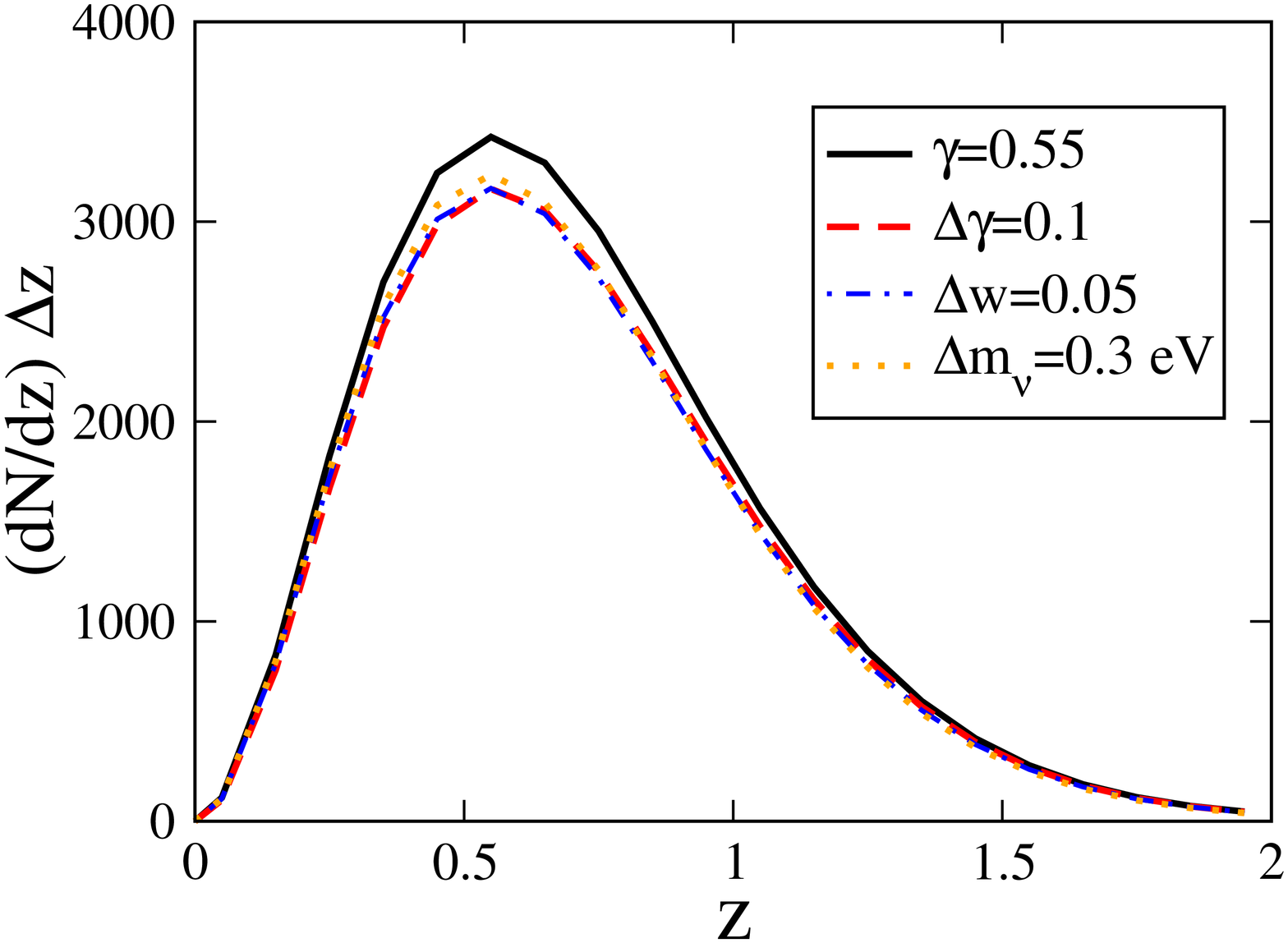, height=3.5in, angle=-90} 
\psfig{file=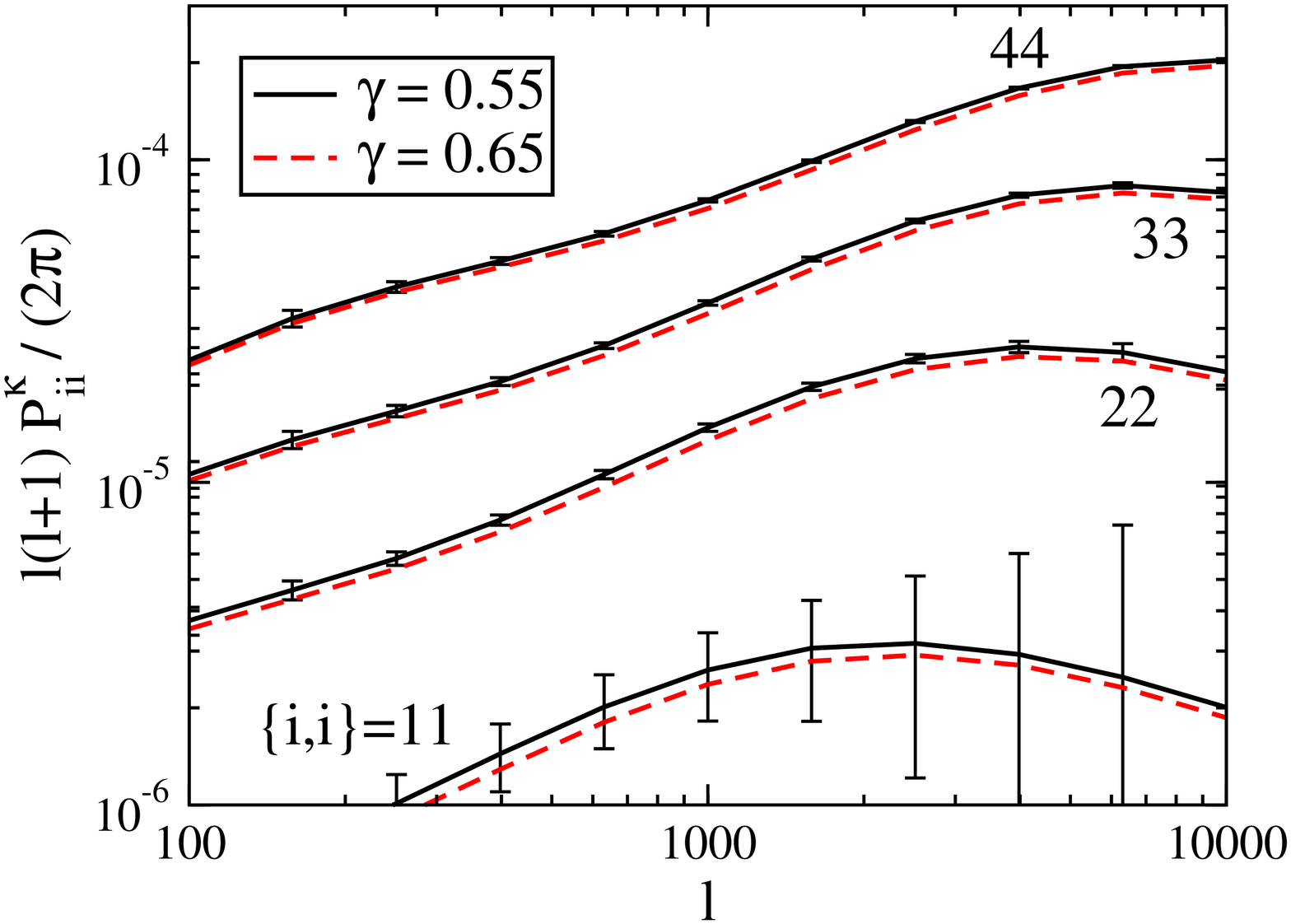, height=3.5in, angle=-90} 
\caption{Left panel: Cluster counts per $\Delta z=0.1$ as a function of
redshift for our fiducial survey, for two values of the growth index
($\gamma=0.55$ and $0.65$).  Also shown are cases when the equation of state
and neutrino mass have been perturbed from their fiducial values by $0.05$ and
$0.3$ eV respectively. Right panel: Auto-correlation power spectra of 4-bin
weak lensing tomography for the two values of the growth index, with
statistical errors shown for the $\gamma=0.55$ model.  
}
\label{fig:sens} 
\end{center} 
\end{figure*}

The gravitational growth index modifies the linear growth factor, which then
propagates into the full nonlinear growth of structure.  Here we examine some
issues related to the nonlinear regime.  In particular, clusters of galaxies
involve nonlinear growth and we might wonder whether special sensitivity to
modified gravity arises in this regime.  The left panel of Fig.~\ref{fig:sens}
shows cluster counts per $\Delta z=0.1$ as a function of redshift for our
fiducial survey (motivated by the South Pole Telescope), for two values of the
growth index ($\gamma=0.55$ and $0.65$). Since we normalized the power spectrum
at high redshift, the number density of clusters is independent of the growth
index at high redshift. As $z\rightarrow 0$, the volume element, which is
independent of the growth index by definition, dominates over the number
density and makes the counts go to zero for either model.  Therefore the
biggest difference between the two models is around the peak of the redshift
distribution at $z \sim 0.6$, as illustrated in the left panel of
Fig.~\ref{fig:sens}. Also shown are number count predictions when the equation
of state and neutrino mass have been perturbed from their fiducial values by
$0.05$ and $0.3$ eV respectively.  While the strong degeneracy of the growth
index with other cosmological parameters is apparent, it is worth noting that
clusters can in principle provide several other observables to break this
degeneracy, chief among them being the mass information.  However, in the
nonlinear regime we are at increased risk of the simple MMG model breaking
down, possibly requiring model dependent simulations for a specific theory of
gravity.

Weak lensing also involves scales in the quasi-linear and nonlinear regimes.
The sensitivity of a weak lensing survey is shown in the right panel of
Fig.~\ref{fig:sens}.  For simplicity, we consider the same fiducial survey as
before but with 4-bin (instead of 10-bin) tomography, with divisions $z=[0,
0.5], [0.5, 1], [1, 1.5], [1.5, 3]$.  We show the four auto-correlation power
spectra with corresponding statistical errors for $\gamma=0.55$, and the same
but without errors for $\gamma=0.65$. The raw signal-to-noise for
distinguishing the two values of the growth index increases with redshift,
giving an advantage to a deeper survey. The angular scale at which the two
values are best distinguished decreases with redshift, so the multipole
increases, going from $\ell\sim 500$ for the first bin to $\ell\sim 5000$ for
the last bin, making advantageous a higher resolution survey.

We have also checked that the sensitivity to the growth index increases as the
density of resolved source galaxies $n_g$ and their mean distance, parametrized
by the mean of the redshift distribution, are increased.  All these factors
indicate that a space-based weak lensing survey has certain definite advantages
for testing beyond Einstein gravity.  At this time, however, we have not
further pursued these sensitivity tests nor tried to devise an optimal strategy
to determine the growth index. The reason is that by far the dominant
uncertainty is a theory uncertainty --- our ability to predict the nonlinear
clustering statistics and associated observables --- in a given modified
gravity theory.

\subsection{Uncertainty and bias} 

So far we have included nonlinear scales when obtaining constraints from 
measurements involving the growth of structure.  Indeed, none of the 
cosmological growth probes is solely a linear theory probe, with the 
possible exception of the
Integrated Sachs-Wolfe effect (which, however, very directly depends on the
metric potentials which themselves are likely to be altered in the modified
gravity theory).   By including nonlinear scales we increase the statistical 
discrimination power with respect to growth but may bias the results as 
a result of employing an improper nonlinear prescription for the modified 
gravity growth.  We now consider the trade off between these two trends. 

\begin{figure}[!t]
\begin{center} 
\psfig{file=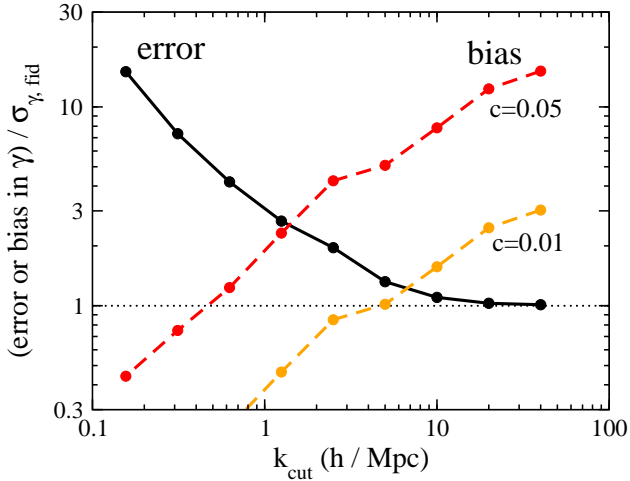, height=3.5in, angle=-90} 
\caption{The solid (black) line shows the degradation in constraints on the
growth index $\gamma$ from a SNAP-type weak lensing survey (plus SN+CMB) as
small-scale weak lensing information is increasingly dropped. For each value of
$k_{\rm cut}$, we drop all information from $k>k_{\rm cut}$ following
\cite{nulling} and compute the ratio of the constraint on $\gamma$ relative to
the case when all information is used, that is, when $k_{\rm cut}=\infty$. The
dashed curves (dark (red) and light (orange)) lines show the bias in the 
growth index divided
by the $k_{\rm cut}=\infty$ statistical error for the bias model given in
Eq.~(\ref{eq:SW_bias}) with $c=0.05$ or $0.01$.  The optimum choice of $k_{\rm
cut}$, with the least risk, occurs near the intersection of the error and bias
curves.  }
\label{fig:kcut} 
\end{center} 
\end{figure}

Since weak lensing probes a range of scales, we can consider limiting the weak
lensing information to scales $k\leq k_{\rm cut}$ and investigate how the
information on the growth index is degraded.  
This ``k-cutting'' can remove those physical scales where we lack 
dependable estimates of modified gravity effects. 
The most straightforward way to implement the
cutting is to use weak lensing power spectrum nulling tomography (see \S5
of \cite{nulling}) where, for a given multipole $\ell$, only the lens planes
with distances $r(z)>\ell/k_{\rm cut}$ are allowed to contribute
information. Higher $\ell$ then contribute increasingly less information, and
exhaust all information below $k_{\rm cut}$. For a fixed $k_{\rm cut}$, we
compute the ratio of the error in the growth index relative to the error with
$k_{\rm cut}=\infty$. The results are shown with the solid line in
Fig.~\ref{fig:kcut}.  Restricting the information to purely linear scales
($k_{\rm cut}<0.2\hmpcinv$) leads to degradations in the marginalized error in
$\gamma$ of more than a factor of 10 relative to the full nonlinear fiducial
case. However, when the quasi-linear scales are used ($k_{\rm cut}<1\hmpcinv$),
the degradation is kept to a factor of a few, and the resulting constraints on
the growth index are still interesting.  Therefore, even if the nonlinear
prediction out to $k\approx 10\hmpcinv$ ($\ell\approx 10000$) in modified
gravity theories is unfeasible, efforts to understand predictions on
quasi-linear scales are well worthwhile.

Since we desire to retain values of $k$ beyond the linear regime, for their
leverage, it is instructive to make at least an approximate estimate of the
bias that might be induced by keeping such data.  Within the halo model, the
linear regime corresponds to the 2-halo term of clustering between dark matter
halos, while the nonlinear regime involves the 1-halo term of the profile and
concentration within a halo.  For $k\approx 0.2-1\hmpcinv$, the 1-halo 
contribution may be approximated by a white noise term in the power spectrum 
\cite{schulzwhite}, due to Poisson fluctuations in the number of halos.  
In other words, 

\begin{equation}
\Delta^2(k, z)\simeq \Delta^2_{\rm lin}(k, z)+\left ({k\over k_*(z)}\right )^3.
\label{eq:SW_term}
\end{equation}

\noindent We find a good fit to the full power spectrum in the 
quasi-linear regime for $k_*(z)=(5/3)\,k_{\rm nl}(z)$, where the nonlinear 
scale $k_{\rm nl}(z)$ is defined via $\Delta^2(k_{\rm nl}(z),z)=1$.  

Since neither the presence of a 1-halo/2-halo split nor the Poisson 
fluctuations in number should rely on the specific gravity 
theory, we adopt this approach and consider what happens if we improperly
estimate the 1-halo effects.  This would shift the white noise term to a
different value (equivalent to changing the scale at which the 2-halo and 
1-halo terms are comparable), i.e.\ biasing the power spectrum by

\begin{equation}
\delta(\Delta^2(k, z)) = c\left ({k\over k_*(z)}\right )^3
\label{eq:SW_bias}
\end{equation}

\noindent where $c$ is a dimensionless constant which represents the fractional
error in the 1-halo term. We then use the Fisher formalism to propagate this
offset into biases on cosmological parameters.  (The expansion parameters $w_0$
and $w_a$ are not appreciably affected by nulling, so we focus on $\gamma$.)

The less of the nonlinear regime we use, the less
effect the misestimated 1-halo, or $c$, term has. The results are shown in
Figure~\ref{fig:kcut} where the dashed curves give the bias in the growth index
divided by the fiducial statistical error (i.e.\ the statistical error for
$k_{\rm cut}=\infty$). It is clear that, as we perform increasingly more
drastic nulling, decreasing the value of $k_{\rm cut}$, the bias/error ratio
decreases significantly.  For $c=0.05$ (0.01), cutting information beyond
$k_{\rm cut}=1$ (6) $\hmpcinv$ gives bias in $\gamma$ that is below the
statistical error, while increasing the error by a factor of three (25\%). 
One might expect the bias model of Eq.~(\ref{eq:SW_bias}) to be cut off 
in the stable clustering regime, causing bias in $\gamma$ to level off 
or decline at $k$ greater than a few times $k_{\rm nl}$, e.g.\ $k>1\hmpcinv$. 
This would allow use of more of the nonlinear regime. 

\subsection{Scale dependent growth}

MMG, with the gravitational growth index formalism, has been adopted as the
simplest reasonable method of accounting for the effects of beyond
Einstein gravity on cosmological probes involving the growth of structure.
One part of its simplicity is that $\gamma$ acts in a scale independent
fashion; this should reproduce global effects such as time varying
gravitational coupling.  However many modifications to gravity will
introduce scale dependence in the growth.  For example, in the DGP
braneworld we might expect changes to the growth equation beyond the
varying coupling on both small and large scales, due to the Vainshtein
radius (where the scalar degrees of freedom become relevant) and
relativistic effects respectively (R.\ Scoccimarro, private
communication).  

Scale dependence on small scales affects the nonlinear regime, and this
is just what we looked at with the bias calculations above.  On large
scales, we may treat the modification of the source term mode by mode
in the linear growth equation.  From a harmonic analysis of metric
perturbations, \cite{jlw1,jlw2} found long wavelength corrections to the
Poisson equation; recall from \S\ref{sec:index} this alters the factor
$Q$ in the source term $Q\delta$.  Such JLW corrections would multiply
$\om(a)$ by $Q\sim 1+\alpha e^{-(k/k_{\rm JLW})^2}$.  We can then either solve
the modified differential equation for the linear growth, or we can
retain the growth index approach, but similarly multiply $\om(a)$ in
Eq.~(\ref{eq:groexp}) by that same factor (this can also be viewed as
making the growth index $\gamma$ a function of $k$).  The characteristic scale
is the horizon scale, $k_{\rm JLW}\sim H$, and the amplitude $\alpha\sim O(1)$.
We find that effects on the growth are then negligible for
$k>10^{-3}\hmpcinv$.

Thus horizon size scale dependence, as should hold for the DGP case and
scalar-tensor theories as well, will have essentially no effect on the weak
lensing probe, as the WL power spectrum errors on the near-horizon scales are very
large due to sample variance; see e.g.\ Fig.~(10) in
\cite{huterer_thesis}. Weak lensing can be reasonably treated by the scale
independent $\gamma$ factor over $10^{-3}\lesssim k/\hmpcinv\lesssim 10$.  Such
scale dependence may however need to be taken into account for attempts to use
the ISW effect to probe cosmology, as mentioned previously.  Only if the
gravitation theory possesses a substantially smaller scale, approaching the
nonlinear scale, as in the phenomenological alterations of the Poisson equation
in \cite{whitekoch,shirata,sealfon,stabenaujain}, are we forced to more
elaborate parameterizations than $\gamma$.

The specific optimum of the trade off between leverage on cosmological
parameter constraints from the added information of smaller scales and
increasing risk of bias from gaps in our understanding will depend on the
specific gravity theory.  Since this is what we are trying to obtain insight
into, a rational, model independent approach might be to carry out both a wide
area survey to squeeze the most statistical power out of the relatively weakly
discriminating low $\ell$ (linear) regime and a deep, high resolution and high
number density survey to probe the richer high $\ell$ (quasi- and nonlinear)
regime.  This indicates possible strong complementarity between a ground-based
weak lensing survey such as LSST \cite{lsst} and a space-based survey like
SNAP.

\section{Discussion}

This article presents an approach for simultaneous fitting of the cosmic 
expansion and the theory of gravity.  
We have advocated a ``minimalist'' strategy of distinguishing modified gravity
from dark energy, which consists in measuring a single parameter, the growth
index $\gamma$. In addition to reproducing the linear growth function for 
essentially all standard gravity, dark energy models (parametrized with 
$w_0$ and $w_a$), the growth index
also fits the linear growth of a single known modified gravity theory, the DGP
braneworld scenario. Therefore, it is reasonable to expect that the growth
index can be used to measure deviations from standard gravity: given the
measurements of the background expansion rate (parameterized, say, by
$\Omega_M$, $w_0$ and $w_a$), standard gravity predicts the value of $\gamma$,
and a statistically significant deviation from this value can in principle be
interpreted as evidence for -- and characterization of -- beyond Einstein 
gravity. 

This is a first step, hence our emphasis in calling it Minimal Modified
Gravity.  One could examine more complex schemes but these may be more model
dependent or
employ more parameters and are therefore likely to give weaker results.  The
MMG approach therefore has more statistical power, being more suitable to
near-future data, while we think the loss in generality is minimal.  We also
emphasize the advantage in retaining the maximal physics information by
treating the expansion effects on growth through the effective equation of
state, giving clear separation from deviations in the gravity theory.

How accurately such a program can reveal beyond Einstein gravity -- dark
physics vs.\ physical darkness -- is the main topic of this paper. We have
shown that measurements from weak gravitational lensing, Type Ia supernovae,
and the CMB combined can measure the growth index to about 8\%, or to $\pm
0.04$ around its $\Lambda$CDM value (and galaxy cluster data could potentially
reduce this even further). At the same time, the constraints on other cosmological
parameters are not appreciably degraded, essentially because the surveys probe
a range of scales and thus their complementarity breaks the degeneracies
between parameters.

One particular concern in the program of distinguishing general relativity from
modified gravity is the nonlinear density regime of structure formation. 
Even for the limited number of
well-defined modified gravity theories, details of nonlinear clustering are
currently unknown. While in principle the nonlinear structure formation is
calculable from N-body simulations of modified gravity, creating these
simulations in practice is extremely difficult except for some very simple
cases. The structure and evolution of galaxy clusters, which are nonlinear
objects, is fairly strongly dependent on the nonlinear physics, and is 
consequently problematic. 

Weak lensing, on the other hand, probes a range of scales, and we studied how
our results behave if we drop small-scale (that is, nonlinear)
information. Using the nulling tomography approach and a reasonably well
motivated toy model for bias due to uncertainty in nonlinear structure, 
we found that cutting out
the small scale information ($k\gtrsim 1\hmpcinv$) can lead to significant
decrease in the resulting bias in the growth index, at the expense of
increasing the statistical error in it by a factor of a few.

Other cosmological observables exist with sensitivity to the growth of
fluctuations, and hence can be used to constrain MMG, but we have
not discussed them in any detail.  For example, the bispectrum of weak
gravitational lensing is a potentially powerful probe \cite{takada_jain};
however, it has proved to be a tough task to calibrate the bispectrum even in
standard general relativity.  The same holds for 
Lyman-alpha forest observations. The Integrated Sachs-Wolfe effect is a 
potentially strong
discriminant of modified gravity models (see in particular recent predictions
of the ISW in DGP models \cite{song_sawicki_hu}); however, the metric
potentials are particularly sensitive to the structure of the modified gravity
theory.  Neither we nor anyone else has yet succeeded in finding a generic
parametrization of the deviations from general relativity for the ISW effect.

Our work outlines a first step in treating modified gravity models. At the time
of this writing, there is hardly a single well-defined modified gravity theory
that does not look like standard general relativity in terms of observables,
but is not already ruled out or disfavored by data. Considerable effort is
underway to construct such theories (e.g.\
\cite{luess2,chiba,carroll_trodden,weller,odintsov,navarro,gabadadze,song,zhang,
amendola,defelice,msg,gauss1,gauss2,ghost}; for a review see
\cite{lue_overview,uzan,trodden_review}) and test them experimentally.

It is heartening that the strong complementarity in cosmological probes 
such as the combination of weak lensing, supernovae, and the CMB does 
provide important information on the question of dark physics vs.\ 
physical darkness.  Such data from next generation experiments, here 
calculated specifically for SNAP and Planck, does furnish a real test 
not just of individual models but of the physics framework beyond 
Einstein.  We will be able to measure the effective equation of state 
describing the cosmic expansion and simultaneously reveal the theory of 
gravity.  By continuing forward with advances in measurements, theory, 
and computation we can lift the darkness on the new physics. 

\section*{Acknowledgments} 

This work has been supported in part by the Director, Office of Science,
Department of Energy under grant DE-AC02-05CH11231 and by NSF Astronomy and
Astrophysics Postdoctoral Fellowship under Grant No.\ 0401066.  We 
gratefully acknowledge discussions with Wayne Hu (who supplied the full 
CMB Fisher matrix) and Martin White.

\end{document}